\documentclass[%
 reprint,
 aps,
 prl,
]{revtex4-2}

\usepackage{graphicx}% Include figure files
\usepackage{dcolumn}% Align table columns on decimal point
\usepackage{bm}
\usepackage{float}
\usepackage{hyperref}
\usepackage{mathtools} 
\usepackage{physics}
\usepackage{subfiles}
\usepackage{appendix} 
\def\energy{E}

\newcommand{\figref}[2]{Fig.~\ref{fig:#1}\,(#2)}

\newcommand{\sm}[1]{{SM }\,(#1)}

\begin{document}

\preprint{APS/123-QED}

\title{Multi-field Return Point Memory}

\author{Nathaniel Croce} 
\affiliation{Department of Mechanical Engineering and Materials Science, Duke University, Durham, NC, 27708, USA}
\author{Hossein Salahshoor} 
\affiliation{Department of Civil and Environmental Engineering, Duke University, Durham, NC, 27708, USA \\
Department of Mechanical Engineering and Materials Science, Duke University, Durham, NC, 27708, USA}
\author{D. Zeb Rocklin}
\affiliation{School of Physics, Georgia Institute of Technology, Atlanta, GA 30332}

\date{\today}

\begin{abstract}
Non-equilibrium systems display memory, a dependence not merely on their present environment but on previously applied fields. Multistable systems such as spin glasses, martensites and granular matter have exponentially many microstates consistent with an applied field, making their rich dynamics difficult to control. Control and order can be achieved through the concept of partial ordering, which we here generalize to systems subject to multiple control fields. We demonstrate, within the model system of the zero-temperature Ising model, that this leads to return-point memory, in which an applied sequence of fields restores the hysteretic system not only to a previous magnetization, but to a previous exact microstate. The multiplicity of fields grants more precise and complex control of the system, with different classes of operations displaying commutative and noncommutative behavior. This grants new insight into how physical systems can remember, learn, and be trained.
\end{abstract}

\iffalse
\section{Things to cite}

\begin{itemize}
    \item Jim Sethna's hysteresis and hierarchies
    \item ``Return point memory in knitted fabrics'' on arXiv
    \item Andrea Liu, Doug Durian and company and Xiaoming Mao's recent stuff on learning in mechanical (and electric) systems
    \item Glaucio Paulino recent noncommutativity operations?~\cite{zhao2025modular}
    \item Sue Coppersmith and Alan Middleton
    \item Jim and Zeb's review articles
    \item Keim, Nagel etc.'s review articles on memory, and Nagel's on soft matter
    \item Karen Daniels' on origami tubes
    \item Sid Nagel, Nate Keim Joey Paulsen's memory in solid materials thing
\end{itemize}
\fi

\maketitle

Physical systems in equilibrium have many commonalities, including the laws of thermodynamics, detailed balance, fluctuation-dissipation theorems and observable physical values derived from phase-space averaging~\cite{pathria2017statistical}.  Nonequilibrium
systems, by contrast, exhibit a much broader range of behaviors, but far fewer
universal organizing principles.  One of the most important such behaviors is
memory, which refers to the ability of a system to encode information about previously applied
perturbations in its present state and future response~\cite{keim2019memory,
Paulsen2025MechanicalMemories}.  
Memory formation has been observed across a wide range of driven and disordered
systems, including charge-density waves~\cite{Coppersmith1987PulseDurationMemory,
Coppersmith1997ShortTermMemories}, granular materials~\cite{Josserand2000GranularMemory},
suspensions and disordered solids~\cite{Keim2011GenericTransientMemory,
Paulsen2014MultipleTransientMemories, Lahini2017NonmonotonicAging,
Dillavou2018FrictionalMemory, Keim2022MechanicalAnnealing,
Shohat2023DissipationMemory}, structural and spin glasses~\cite{
Fiocco2014EncodingMemory, Mukherji2019MechanicalMemoriesYield,binder1986spin},
and active matter~\cite{marchetti2013hydrodynamics}.  More broadly, memory is
central to function in biological systems ~\cite{Hopfield1982NeuralNetworks,Amit1985SpinGlassNeuralNetworks}, and newly emerging physical learning in tunable mechanical networks~\cite{
Stern2020ContinualLearning,stern2021supervised,dillavou2022demonstration,dillavou2024machine,stern2023learning,li2024training,mandal2024learning}. These examples suggest that memory converts a
disordered or multistable material from a passive medium into an information-bearing physical system.

Among the many forms of material memory, return-point memory occupies a special
place because it is exact and reproducible.  In a system with return-point
memory, a bounded excursion of an applied field returns the system not only to a
previous macroscopic observable, but to a previous microscopic state.  The
canonical theoretical setting is the zero-temperature random-field Ising model
driven by a single scalar field, where ferromagnetic interactions, metastable
relaxation, partial ordering, and the no-passing property imply nested
hysteresis loops and exact return to prior states~\cite{
middleton1992asymptotic,sethna1993hysteresis,sethna2006random,
Deutsch2004ReturnRPM}.  This framework has shaped our understanding of
disorder-driven hysteresis and avalanches in magnets, martensites, and other
irreversible systems~\cite{sethna2017deformation}.  However, the classical
theory relies crucially on a scalar driving where histories can be ordered by their maxima and minima along a single field. Yet, many programmable and mechanical systems of current interest are controlled by
more than one independent input, such as multiple forces, displacements, stresses, strains,
actuation patterns, fields, or boundary motions.  In such systems, the applied
drive is a path in a multidimensional control space, and there is no natural
total ordering of histories.  This raises a basic question: can return-point
memory survive under multi-field or vector-valued driving, and if so, what replaces the scalar
ordering structure that makes the classical theory work?

In this Letter, we introduce the multi-field random-field Ising model as a minimal
setting for addressing this question.  Each spin is coupled ferromagnetically to
its neighbors and experiences quenched disorder, but the external drive consists
of multiple independently controlled fields acting on different spatial patterns
of spins.  We show that monotone paths in this multidimensional control space
lead to the same final microstate, independent of the particular path taken.
For non-monotone paths, however, distinct field operations generally do not
commute: applying the same changes in different orders can lead to different
final microstates.  Despite this noncommutativity, we prove and demonstrate a
multi-field form of return-point memory in which bounded variations in control
space exactly restore the system to an earlier  microstate.  This extends
return-point memory from scalar hysteresis loops to multidimensional protocols
and suggests new principles for organizing, retrieving, and transforming
microstates in trainable multistable matter.

\begin{figure}
  \centering
  \includegraphics[width=0.5\textwidth]{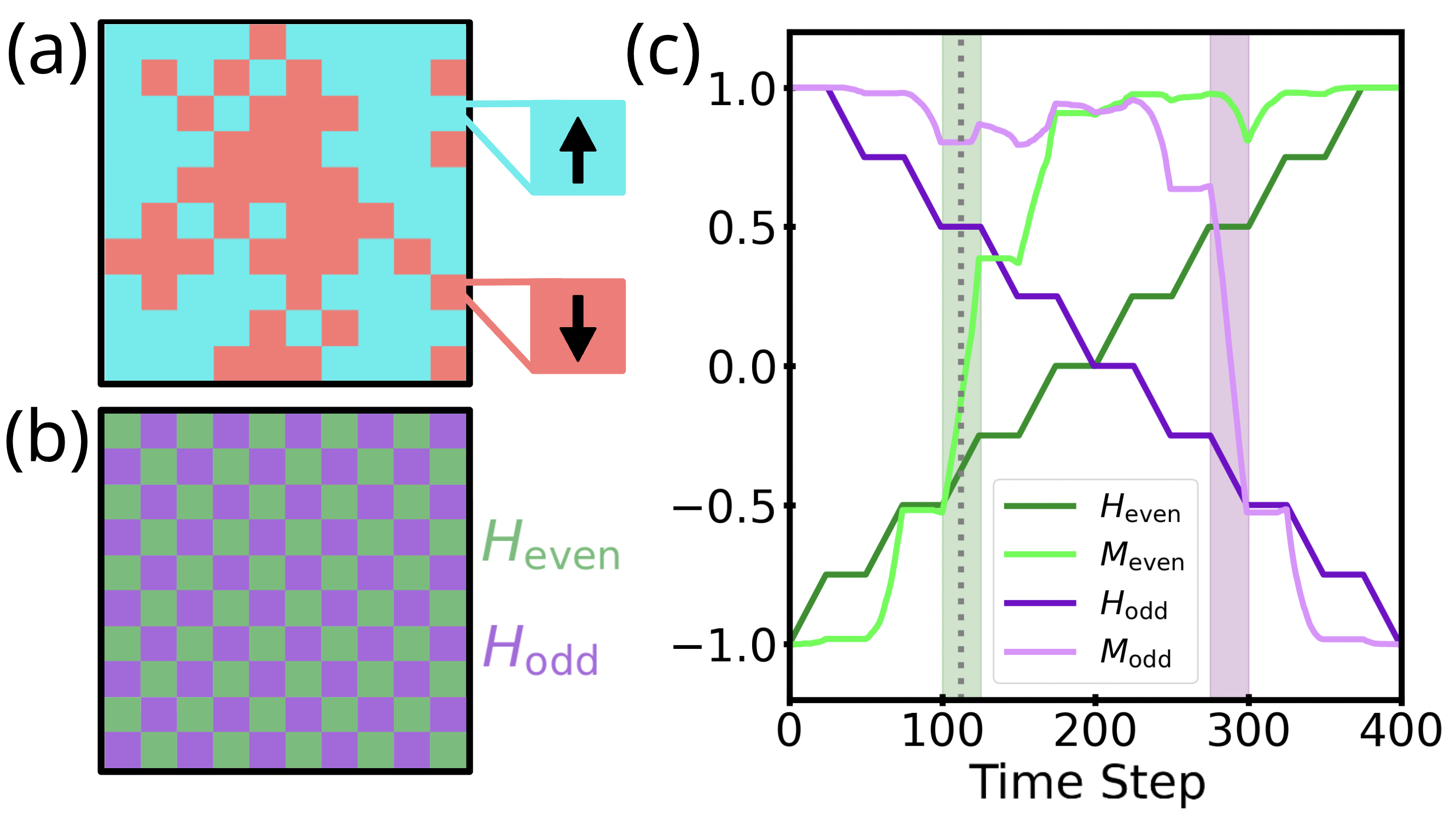}
  \caption{\textbf{Application and behavior of system with two external fields.} (a) An arbitrary snapshot of a $10\times 10$ subset of a $100\times 100$ spin configuration at time $t=112$: light blue squares represent a positive spin at the site (+1) and light red squares represent a negative spin (-1). (b) The $H_{\text{even}}$ (green) and $H_{\text{odd}}$ (purple) fields are applied in a checkerboard pattern across the $10\times 10$ lattice of spins. (c) Time evolution of the external fields $H_{\text{even}}$ (dark green) and $H_{\text{odd}}$ (dark purple) over time, and the corresponding magnetization at the even sites $M_{\text{even}}$ (light green) and the odd sites $M_{\text{odd}}$ (light purple) when $H_{\text{even}}$ and $H_{\text{odd}}$ are applied to a $100\times 100$ lattice of spins in the checkerboard pattern demonstrated in (b). The shaded region corresponds to instances where one of the two H fields are held constant. Dashed-line corresponds to arbitrary time step $t=112$. 
    }
  \label{fig:setup}
  \vspace{-10pt}
\end{figure}

We consider a system based on the Ising model~\cite{ising1925beitrag}, which has seen widespread application in statistical physics~\cite{sethna2021statistical,cardy1996scaling} consisting of a square grid (periodic boundary conditions are assumed) of spins, each of which can be either up (+1) or down (-1), as shown in \figref{setup}{a} and the associated Supplementary Video. We retain the Ising model's original language of magnetic spins and fields, but the same model has been effectively applied beyond standard statistical physics systems to neurons~\cite{roudi2009ising}, buckling beams~\cite{kang2014complex} and curving surfaces~\cite{plummer2020buckling,plummer2022curvature}. We term our model the multi-field Ising model (MFIM), and define it by the following energy functional: 

\begin{align}
\energy = - J\sum_{\langle i,j\rangle} s_i s_j - \sum_{i} \left(h_i^0 + \sum_a c_a(t) H_{a i}\right) s_i.
    \label{eq:energy}
\end{align}

\noindent Here, $J>0$ is a coupling constant between each spin and its four neighbors and $h_i^0$ is a quenched (static) field that is different at different sites, where we typically take $h_i^0$ to be independently and identically normally distributed with mean zero and standard deviation $\sigma$. We introduce the concept of multiple \emph{control fields} $c_a(t)$ which each undergo some prescribed evolution under time $t$. The coefficients $H_{ai} \ge 0$ describe how strong a field control $a$ exerts on site $i$. Note that the total energy can be expressed in the form $E=-\sum_i h_i s_i$, where the local field  $h_i$ is a combination of interactions, quenched disorder and the control fields.

Appropriate for macroscale systems, we take our system to be at zero temperature, meaning that after a change in the external fields, spins flip to lower the energy, eventually reaching a local minimum in the multistable energy landscape induced by disorder. As shown in the Supplementary Material [\sm{IV}], the order in which spins flip does not affect the final state. We can define a \emph{partial ordering} on the system states, such that one state is greater than or equal to another if each spin in the first state is greater than or equal to the corresponding spin in the second state. While the vast majority of pairs of microstates are not ordered in this sense, the states obey a \emph{no-passing condition}, in which two ordered states maintain their order when subjected to the same fields. This type of condition was first identified for sliding charge-density waves~\cite{middleton1992asymptotic} and later extended to spin systems~\cite{sethna1993hysteresis}.

 The special case of only a single field that applies equally to all spins ($H_{1i} = \textrm{const.}$) has been extensively studied and referred to as the random-field Ising model (RFIM)~\cite{sethna1993hysteresis,fytas2018review, belanger1991random,sethna2006random}. In this Letter, we mainly focus on the case of checkerboard fields, in which one field is applied to all even sites (in which their horizontal and vertical indices sum to an even integer) and another field is applied to all odd sites, as shown in \figref{setup}{b},  although we also include an example of many-field control. In the particular control protocol shown in \figref{setup}{c}, the field on the even sites (in units such that $J=1$) is ramped up, resulting in some of the spins to which this field is applied flipping up. This in turn can flip some of the other spins up, so that both even and odd magnetizations (defined as the average values of the corresponding spins) increase. In other time steps, the odd field is decreased, resulting in both even and odd spins flipping down, so that the overall magnetization  is roughly maintained while shifting from odd to even magnetization.

\begin{figure}
  \centering
  \includegraphics[width=0.5\textwidth]{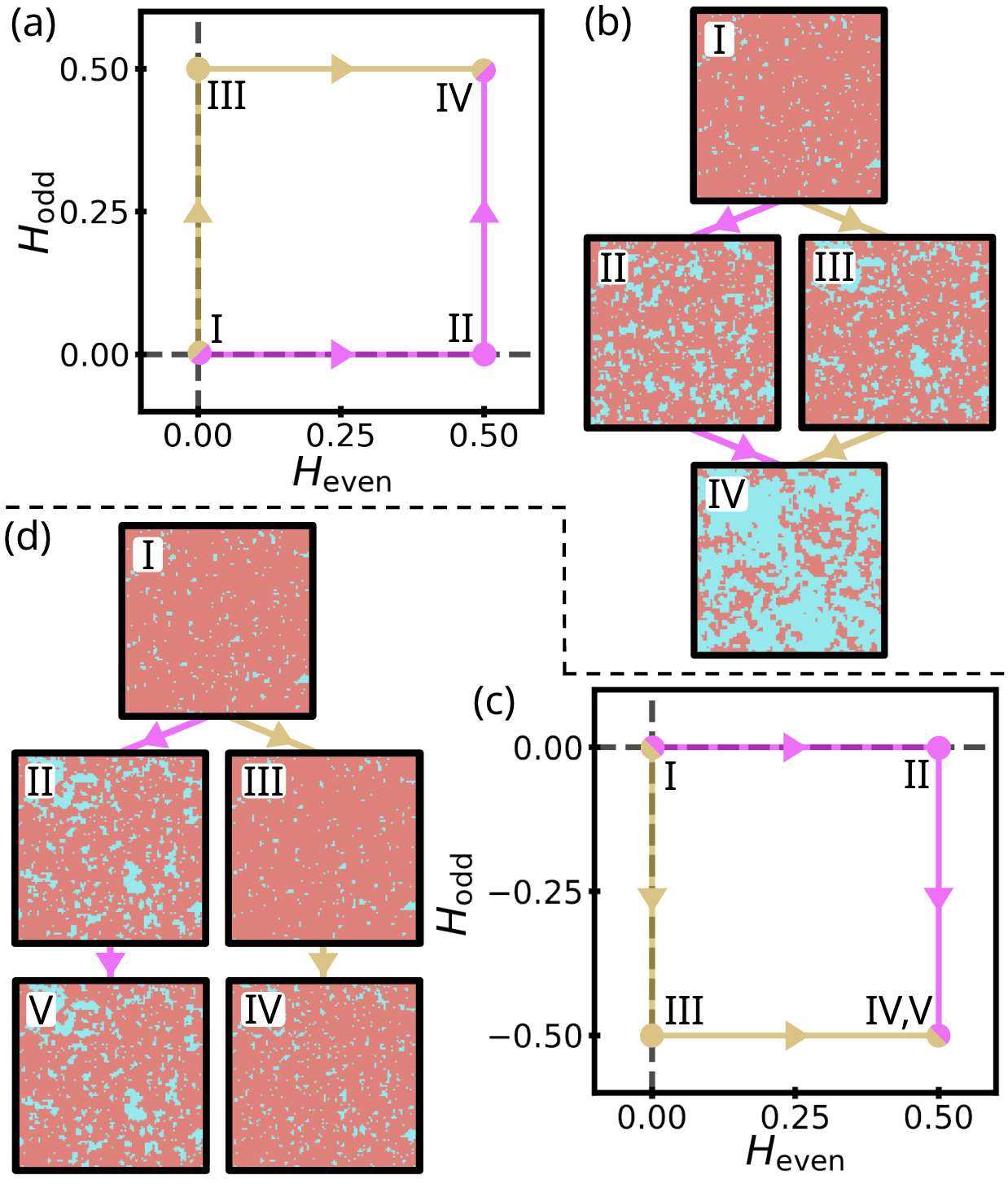}
  \caption{\textbf{Commutativity and non-commutativity in a two field system.} (a) Two systems begin from identical initial configurations (I) and evolve under distinct monotonic fields trajectories in the ($H_{\text{even}}$, $H_{\text{odd}}$) plane. Path 1 (pink) first increases $H_{\text{even}}$, and then increases $H_{\text{odd}}$, while Path 2 (yellow) reverses the order. Because both field components vary monotonically and reach the same final values, the resulting final configurations are identical (IV) as depicted in (b). The intermediate configurations (II,III) differ and exhibit no relation. (c) depicts a similar setup to (a) starting with two identical systems, but they now evolve under distinct non-monotonic field trajectories. Path 1 (pink) first increases $H_{\text{even}}$, and then decreases $H_{\text{odd}}$, while Path 2 (yellow) reverses the order. Although both paths terminate at the same external field values, their final spin configurations (V,IV) differ as depicted in (d). Notably, V is greater than IV because Path 1 begins with an increase while Path 2 begins with a decrease; for the same reason, II is greater than III.
    }
  \label{fig:comnoncom}
\end{figure}

\begin{figure*}
  \centering
  \includegraphics[width=\textwidth]{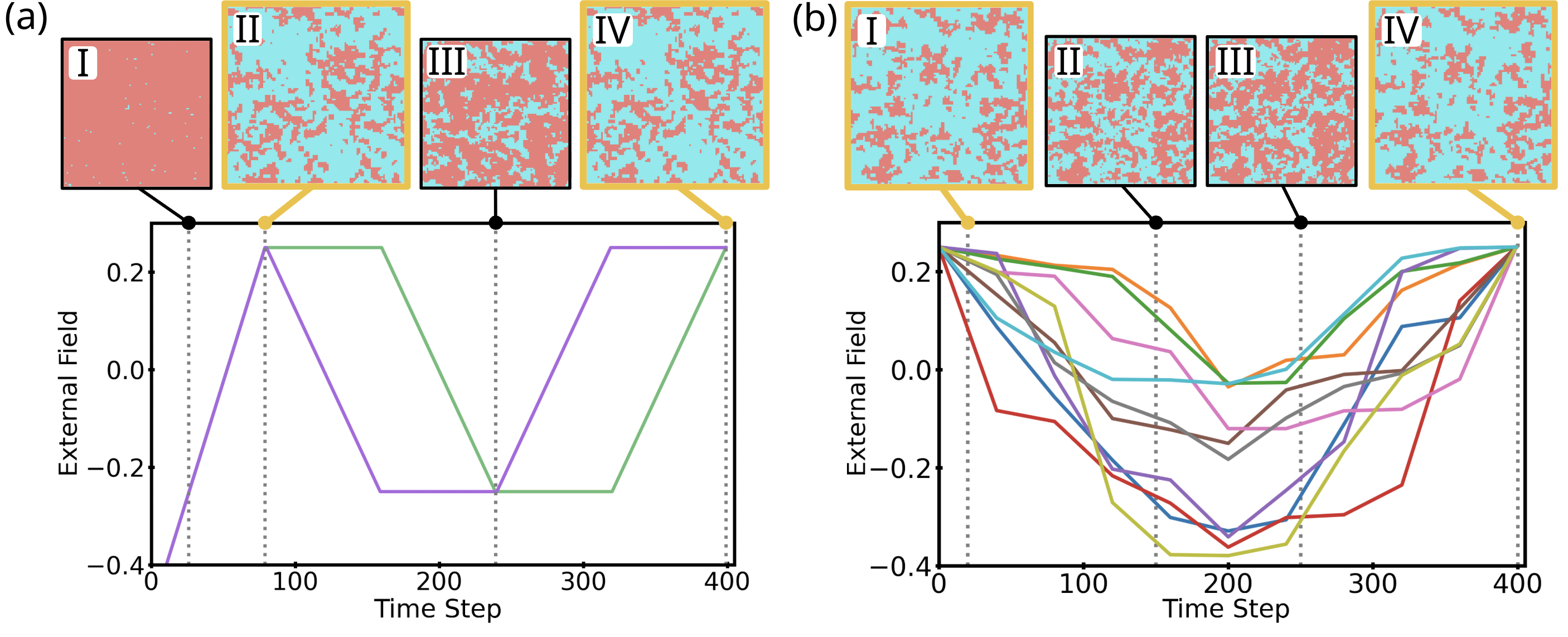}
  \caption{\textbf{Demonstration of return point memory in two-field and multi-field system.} (a) The external fields $H_{\text{even}}$ (green) and $H_{\text{odd}}$ (purple) fields evolve over time as shown. Both fields begin at large negative values and rapidly increase to pass through spin configuration I and reach spin configuration II. The fields then vary independently over time until reconverging to produce final configuration IV, which is identical to  state II. Notice that $H_{\text{even}}$ and $H_{\text{odd}}$  never exceed the extrema reached during state I, consistent with the no-passing rule, and thus states II and III are less than I. (b) $100\times 100$ external fields evolve over time (only 10 fields are depicted on the plot) and are applied to a $100\times 100$  spin lattice, allowing each field to act on only a single spin. The spin configurations exhibit the same return point memory effect demonstrated in (a).
    }
  \label{fig:rpm}
\end{figure*}

In a separate protocol depicted in \figref{comnoncom}{a} (magenta line), we begin in a state with zero field applied, increase the first field, and then subsequently increase the second field. As spins flip to align with the external fields and each other, the system passes through microstates labeled I, II and IV as shown in \figref{comnoncom}{b} (see also the associated Supplementary Video).
 Separately, we consider a protocol (yellow line) in which the increase in the odd field occurs before the increase in the even field. While the system passes through a different set of intermediate states, such as III, than in the previous protocol, it arrives at exactly the same final state (IV).
 This is a demonstration of the more general phenomenon of \emph{equivalence of monotonic paths}. That is, as proven in the Supplementary Material [\sm{VI}], when an initial state is subject to an initial set of fields at some time $t_0$ and evolved monotonically under changing fields to some final state at time $t_f$, the result depends on the initial state and the initial and final fields, but is independent of the particular monotonic path. Here, we say that a path is monotonically increasing only if for every $t_0 \le t_1 \le t_2 \le t_f$ that $h_i(t_1) \le h_i(t_2)$ for each site $i$. Given our positive coefficients $H_{ai}$, this is ensured so long as each control field $c_a(t)$ is increasing. Unlike the single-field result~\cite{sethna1993hysteresis}, this equivalence is over an infinite number of monotonic paths through control space. Because increasing fields can be applied to different sites in any order, these actions can be interpreted as commuting operators acting on the system, akin to the abelian nature~\cite{dhar1990self} of the sandpile model~\cite{bak1987self}.

We  also consider non-monotonic paths, such as the two protocols shown in \figref{comnoncom}{c}. In the first protocol (magenta) the even field is increased and then the odd field is decreased. In the second protocol (yellow) the order of these operations is reversed, so that the system follows a different path in field space to the same final applied control fields.
In contrast with the strictly increasing paths, not only are the respective intermediate states shown in~ \figref{comnoncom}{d} (II, III) different, the final states (V, IV) differ as well (see also the associated Supplementary Video). 
Furthermore, we can say that the protocol in which the field increase occurs earlier generates a final state that is greater than or equal to (in the partial ordering sense) the state resulting from the protocol in which the field decrease occurs earlier, as shown in the Supplementary Material [\sm{VII}]. This sensitivity to the order of the applied fields demonstrates the general \emph{non-commutativity} of non-monotonic control fields. Non-commutativity is a hallmark of history-dependent systems, and is essential to performing logical operations
 and achieving robotic control in multistable systems~\cite{zhao2025modular}.

In general, a multistable system might have many different  states consistent with an applied field, as is the case for frustrated antiferromagnets and spin glasses~\cite{binder1986spin}. Consequently, when the applied fields are evolved over time in a closed loop, it is not generally the case that the system state will similarly return to its initial microstate.  
However, as shown in the Supplementary Material [\sm{IX}], the independence of monotonic paths can be extended to an independence of bounded paths (ones in which the applied fields remain within a given range of values), leading to a novel form of \emph{return-point memory}~\cite{sethna1993hysteresis,sethna2017deformation,Paulsen2025MechanicalMemories}, in which the return to the exact microstate, among hundreds or thousands of stable states that exist at these fields, is achieved. This can be seen in \figref{rpm}{a}, in which both fields are gradually increased, passing through state I, before reaching  a maximum value, at which point the system reaches state II.
The two control fields are then reduced independently. At an intermediate timestep, they reach the same external control fields, but the resulting state, III, lies strictly above state I, with many spins flipped. This is a form of hysteresis, in which the same applied fields result in different states due to history-dependence.
However, as the fields are separately restored to their maximum values, the system is likewise restored to its same microstate, IV, which is identical to state II. The distinction is that between states II and IV, the applied fields remained bounded, lying between the maximum associated with state II and the minimum at initialization---in contrast, between states I and III, the applied fields increased to new maxima.
In \figref{rpm}{b}, the protocol is repeated with 10,000 different fields, one applied to each spin, taken through a random path in field space (only ten are depicted in the graph). Despite the extreme complexity of the protocol, the system is likewise restored to the exact microstate when the fields are brought back to their maximum values, as also shown in the accompanying Supplementary Video.

\begin{figure}
  \centering
  \includegraphics[width=0.5\textwidth]{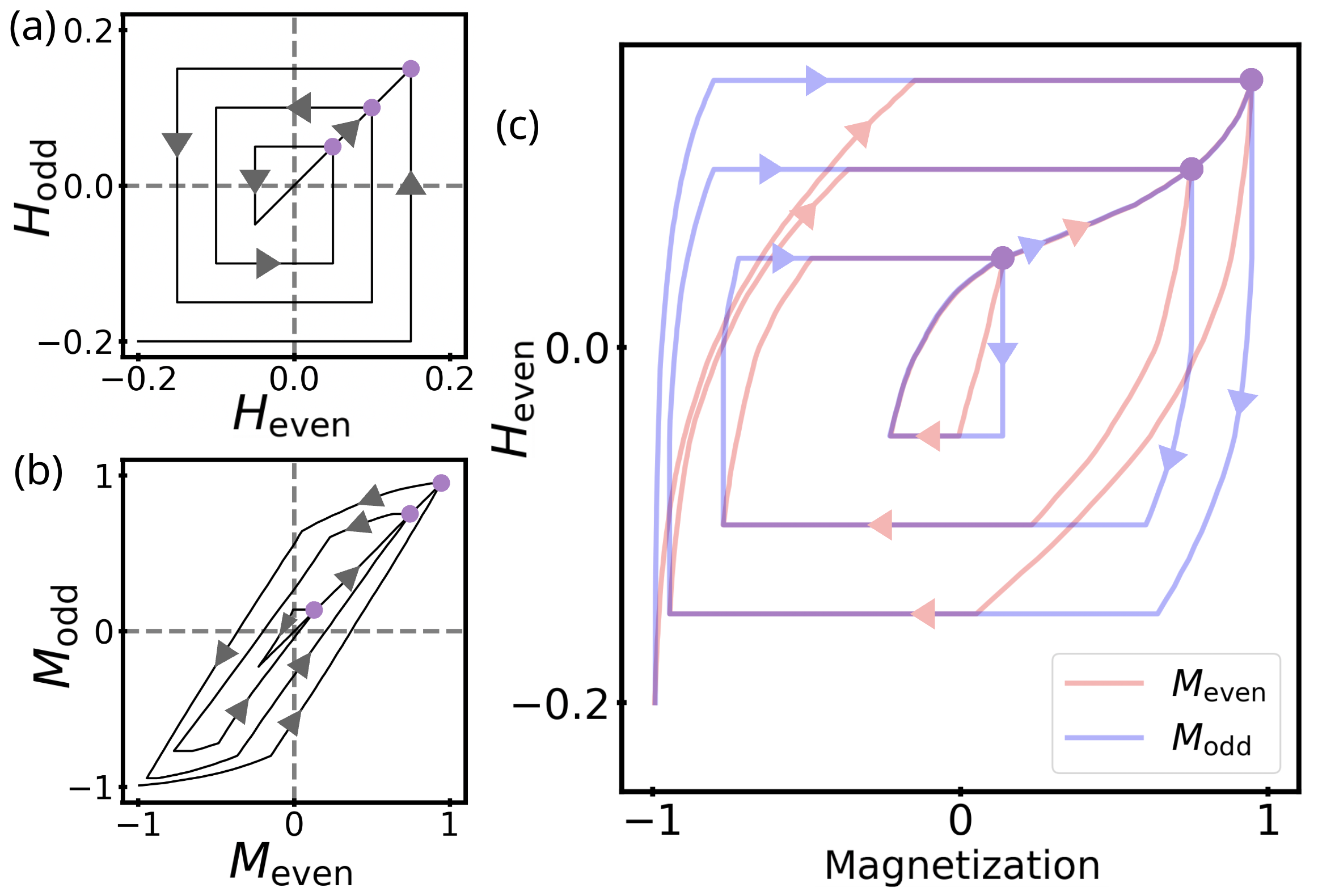}
  \caption{\textbf{Repeated return-point memory effect via spiral control dynamics.} (a) A $200\times200$ spin system evolves under a spiral trajectory plotted in the ($H_{\text{even}}$, $H_{\text{odd}}$) plane. (b) A resulting similar behaving spiral occurs as the system evolves when plotting the magnetization at the even sites and odd sites of the system in the ($M_{\text{even}}$, $M_{\text{odd}}$) plane. (c) The behavior of $M_{\text{even}}$ and $M_{\text{odd}}$ are each demonstrated individually when plotted against $H_{\text{even}}$. The three purple points demonstrate instances of return point memory in the system between when the path first intersects a point during the inward spiral and when the point is intersected again during the final increase. Each instance of return point memory in (c) also has a corresponding instance in the ($H_{\text{even}}$, $H_{\text{odd}}$) plane in (a) and the ($M_{\text{even}}$, $M_{\text{odd}}$) plane in (b).
    }
  \label{fig:rpmloop}
\end{figure}

In \figref{rpmloop}{a}, a two-field protocol is shown in which the fields are ramped up to their maximum values separately, before spiraling up and down through a range of values. Because of the spiral nature of the field path, ramping up the fields results in returning three separate times to previous field strengths (purple dots). As shown in \figref{rpmloop}{b}, this results in returning to the exact same microstates and hence magnetizations (purple dots). \figref{rpmloop}{c} shows the relationship between one of the applied fields and the magnetizations in the same path, as also depicted in the associated Supplementary Video.

We have shown how physical systems subject to complex controls display return-point memory. This enhanced control raises the possibility of developing new applications of the effect, such as proposed magnetic memory systems~\cite{perkovic1997improved},  of training the systems to desired response profiles, or realizing the effect at other length scales or via other types of interactions. Extending the proof of return-point memory, we also advance the possibility of explaining the effect in other systems in which it has been reported, such as antiferromagnets~\cite{Deutsch2004ReturnRPM, shohat2022memory}.

\paragraph*{Acknowledgments}
We gratefully acknowledge 
financial support from the National Science Foundation CAREER program 
(\#2338492), from the Army Research Office through the MURI program 
(\#W911NF2210219) and through NSF PHY2309135 to the Kavli Institute for Theoretical Physics
(KITP) (DZR). 

\paragraph*{Codes}
The computer code used for the generation of data, as well as the videos can be accessed at \url{https://github.com/nateprcroce/MultiField-RPM}.

\nocite{*}
\bibliography{citations}
% \clearpage
% \section{Supplementary Material}
% \subfile{supplementary_material_content}
\end{document}

\preprint{APS/123-QED}

\title{Supplementary Material for: Multi-field Return Point Memory}

\maketitle
%\paragraph*{Introduction}

\section{Model Definition}

As noted in the main text, we
consider a system consisting of a square grid (periodic boundary conditions are assumed) of spins, each of which can be either up (+1) or down (-1). The potential energy of a configuration is of the form:

\begin{align}
\energy = - J\sum_{\langle i,j\rangle} s_i s_j - \sum_{i} \left(h_i^0 + \sum_a c_a(t) H_{a i}\right) s_i.
    \label{eq:energy}
\end{align}

\noindent Here, $J>0$ is a coupling constant between each spin and its four neighbors and $h_i^0$ is a quenched (static) field that is different at different sites, where we typically take $h_i^0$ to be independently and identically normally distributed with mean zero and standard deviation $\sigma$.  $c_a(t)$ are \emph{control fields} which each undergo some prescribed evolution under time $t$. The coefficients $H_{ai} \ge 0$ describe how strong a field control $a$ exerts on site $i$. Note that the total energy can be expressed in the form $E=-\sum_i h_i s_i$, where the local field  $h_i$ is a combination of interactions, quenched disorder and the control fields. Appropriate for macroscale systems, we take our system to be at zero temperature, meaning that after a change in the external fields, spins flip to lower the energy, eventually reaching a local minimum in the multistable energy landscape induced by disorder. As shown below, the order in which spins flip does not affect the final state.

\section{Simulation}

We simulate a two-dimensional $L \times L$ square grid system, with periodic boundary conditions.
The quenched local fields $h_i^0$ from Eq.~(1) are randomly drawn independently and identically from a normal distribution with mean zero and a standard deviation that we choose to be exactly twice the coupling coefficient $J$. 
We initialize the system with all spins pointing down, $s_i = -1$. From there, we apply some control fields $c_a$ and allow the system to relax. In the next time step, we adjust the control fields monotonically (meaning that in a single timestep, either all controls increase or remain constant, or all controls decrease or remain constant). We then allow the system to relax to the local energy minimum before recording its state.

Our particular relaxation protocol is as follows. In an increasing time step, we check each down spin to see whether its local field $h_i$ (combining the quenched field, interactions with its neighbors, and the control fields) is now positive. For each down-spin with a positive local field, we flip it up. However, these spin flips will, in general, lead other local fields to become positive. Consequently, we repeat this procedure, continuing to flip spins up, until no down-spin has a positive local field. We then say that the system has relaxed. In our particular implementation, we traverse the grid flipping spins from left to right and top to bottom. However, different implementations result in exactly the same relaxed state. We have implemented this scheme on a $100\times100$ grid most frequently with two control fields, but also with a single control field and with as many as 10,000 control fields, one for each spin.

\section{Partial ordering}

 We can define a \emph{partial ordering} on the system states, such that one state is greater than or equal to another state if each spin in the first state is greater than or equal to the corresponding spin in the second state. That is, if $s^A$ is a state and $s^A_i$ is the value $\pm 1$ of spin $i$ in state $s^A$, then by definition we say $s^A \ge s^B$ or ``$s^A$ is greater than or equal to $s^B$ if and only if $s^A_i \ge s^B_i$ for each spin $i$, with the analogous definition for $s^A \le s^B$. We say that a set of states is ordered if each state in the set is either greater than or equal to or less than or equal to each other state. However, most pairs of states for large systems are not ordered. For example, consider the states $(+++), (+-+), (++-)$. The first state is greater than or equal to all of the states (including itself), but the second state is neither greater than or equal to or less than or equal to the third state.

 These relationships have three important properties. The first is reflexivity: $s^A \ge s^A$ (every state is greater than or equal to itself). The second is transitivity: $s^A \ge s^B$ and $s^B \ge s^C$ together imply $s^A \ge s^C$ (if a state is greater than or equal to a second state that is itself greater than or equal to a third, then the first state is also greater than or equal to the third). The last is antisymmetry: $s^A \ge s^B$ and $s^B \ge s^A$ together imply $s^A = s^B$ (the only state that is both greater than or equal to and less than or equal to another state is the state itself).

 \section{Algorithmic independence}

 As discussed above, our protocol requires flipping spins to align with their local fields. Here, we show that the state that the system relaxes to is independent of the particular algorithm used.

 For definiteness, consider a time step in which all fields have been either increased or left unchanged. The same logic applies to a time step in which all fields have been either decreased or left unchanged. In this Letter, we only consider individual timesteps in which all fields increase (or remain the same) or all fields decrease (or remain the same), rather than timesteps in which some fields increase and others decrease.

 After the aforementioned increase in field, there are three types of spins: up spins ($s_i = +1$), down spins that are eligible to be flipped up and down spins that are not eligible. Physically, the origin of this ``eligibility'' is that a down spin has a positive local field, generated by a combination of its quenched random field, the interactions from its neighboring spins and the recently increased control fields. Consequently, the eligibility of a down spin to be flipped up is determined entirely by which spins have already been flipped up. In particular, it is independent of the algorithm that determines the order in which spins flip up during relaxation. Note also that flipping spins up only increases the eligibility of other spins to flip up, it never prevents other spins from flipping up. Mathematically, we can say that if $\{s_i\}_2 \ge \{s_i\}_1$ (in the partial ordering sense) then if $s_j$ is eligible to flip up under state 1, it is also eligible under state 2.

 An algorithm is a procedure, deterministic or random, for deciding the order in which to flip eligible spins up. Flipping a spin up can generate new eligible spins, and the algorithm will not terminate until there are no eligible spins. Different potential algorithms include: flip the eligible spin with the greatest local field, flip an eligible spin at random, flip all eligible spins at once, or flip eligible spins in order based on their position. 

 Consider two algorithms, which we will label $a_1, a_2$, each applied to an initial state with some set $\{s_i\}_e$ of spins that are eligible to flip up. Suppose for the purposes of contradiction that these two algorithms lead to different final states. If so, we can say that at least one of the two algorithms (without loss of generality, we can say this is $a_2$) flips at least one spin that the other doesn't. Consider, then, the state of the system under $a_2$ immediately before it flips $s_*$, which we define as the first spin that $a_1$ never flips. In that case, it has flipped some spins which we label $\{s_i\}_m$. By assumption, under $a_1$ these same spins will also flip, though this may come later in the process for this algorithm. That is to say that, by the time that $a_1$ terminates, it will also have flipped all of $\{s_i\}_m$ (as well as, perhaps, some additional spins). Thus, by the time $a_1$ halts, its state must be greater than or equal to (in the partial ordering sense) the state in which $a_2$ flipped $s_*$.

 This implies, under our earlier dynamics, that $s_*$ must become eligible to flip under $a_1$, which also means that it will eventually flip. Thus, we find that the first spin that $a_2$ flips that $a_1$ wouldn't flip cannot exist---$a_1$ must indeed flip it. Because the result is fully general, it follows that any spin that flips under one valid algorithm also flips under any other valid algorithm, so that the relaxed state is independent of choice of algorithm.

\iffalse
 After the fields are increased, some number of spins will be in spin down states ($s_i = -1$) but have positive local fields ($h_i >0$). By assumption, all those spins must flip up before the system is in a relaxed state. We will assume in our analysis that a nonzero number of spins flip during this timestep---if no spins are ever flipped then of course the algorithm cannot flip a different sets of spins.

 Consider two algorithms, and suppose for the purposes of obtaining a contradiction that the first algorithm flips a spin that the second algorithm does not flip. Consider the first point in the first algorithm at which it flips a spin that the second algorithm does not flip, which we will call the divergence point. Now consider the second algorithm. It will, by assumption eventually flip all of the spins that the first algorithm did before the divergence point (and perhaps more as well). Because the second algorithm has flipped all of the spins that the first algorithm had at the divergence point and any additional spin flips it performed can only increase the local fields, we conclude that the local field for the spin that flipped at the divergence point must be positive, and so this same spin will be flipped by the second algorithm as well. Thus, we find that this spin will actually be flipped under both algorithms, and that it is in fact impossible for the algorithms to lead to different relaxed states.
 \fi

 \section{No-passing condition}

 Suppose that we have a system (with fixed local disordered fields $h_i^0$) prepared in two different states that obey a partial ordering, $s^A \ge s^B$, that is a relaxed state of the system with applied control fields $c_a$. Suppose now that we increase some of the control fields and allow the system to relax in both instances. The resulting final states must obey the same ordering as the initial states. The reason for this is that the local fields $h_i$,  include the quenched disorder and controls (identical in the two cases) and the interactions, so that $h_i^A \ge h_i^B$. If we adopt a protocol in which we repeatedly flip all down spins with positive local fields, we see that during each such flipping event $h_i^A \ge h_i^B$, and so in the final, relaxed state, every spin that flipped up in the $B$ case must also have flipped up in the $A$ case. As shown in the previous section, this result must also pertain even if another algorithm is used. It also follows trivially that during this time step in which control fields are increased, the final states of $A$ and $B$ respectively must be greater than or equal to their respective initial states.

By the same logic as the increasing timestep, the ordering must also be maintained if some of the control fields are turned down. We do not consider single time steps in which some control fields are turned down and others are turned up. By composing multiple changes to the controls, we can continue to evolve the system through multiple states while maintaining the ordering. This is known as the no-passing condition, because if one state begins ``behind'' another state, it cannot pass the state (though it can catch up, as we will see).

\section{Equivalence of bounded paths}

Suppose now that we now pursue a protocol of the following type. An initial relaxed state is subjected to controls $c_a^{(1)}$ and allowed to relax. It is then subjected to controls $c_a^{(2)}$ and allowed to relax, and this is repeated for $N$ time steps. Suppose also that the controls are bounded, such that the first value of the controls is the lowest and the last is the highest, for each control $a$:

\begin{align}
    c_a^{(1)} \le c_a^{(n)}  \le c_a^{(N)}, \textrm{ for all } n,a.
\end{align}

\noindent Note that although the controls are bounded they can be nonmonotonic: $c_a^{(n)} \ge c_a^{(n+1)}$ is permitted. Let us denote the state that results from applying this series of fields to an initial state $s$ by the functional relationship $f(s,\{c_a^{(n)}\})$.

We now observe that increasing one or more controls can only increase (or leave unchanged) the state that results immediately after the controls are applied. However, increasing the initial \emph{state} can also only increase (or leave unchanged) the state that results after fields are applied. Consequently, increasing any field during any timestep will increase (or leave unchanged) the final state that results after the $N^\textrm{th}$ timestep.

Now let us consider an alternative protocol, in which the controls are maintained at the minimum value $c_a^{(1)})$ for the first $N-1$ time steps, and only increased to the maximum value of $c_a^{(N)})$ in the final timestep. Because the controls in this second protocol lie at or below the controls in the first protocol in every time step, the final state must also lie at or below the final state resulting from the first protocol. However, because the system relaxes under a set of control fields, applying the same control fields multiple times does not change the resulting state, and the state resulting from the second protocol is merely that resulting from applying the first and last states:

\begin{align}
    f(s,\{c_a^{(n)}\}) &\le f(s,\{c_a^{(1)},c_a^{(1)},\ldots c_a^{(1)},c_a^{(N)}\}) 
    \\
    &= f(s,\{c_a^{(1)},c_a^{(N)}\}).
\end{align}

However, we can apply exactly the same logic to a third protocol, in which the controls are increased to their maximum value in the second timestep:

\begin{align}
    f(s,\{c_a^{(n)}\}) &\ge f(s,\{c_a^{(1)},c_a^{(N)},c_a^{(N)}\ldots c_a^{(N)}\}) 
    \\
    &= f(s,\{c_a^{(1)},c_a^{(N)}\}).
\end{align}

We have thus shown that our original protocol results in a state that is both greater than or equal to and less than or equal the same state. This means that it is equal to this state:

\begin{align}
        f(s,\{c_a^{(n)}\}) = f(s,\{c_a^{(1)},c_a^{(N)}\}).
\end{align}

That is, applying a series of controls, in which the initial control is the lowest value (for each control field separately) and the final control is the highest value, results in a state that is independent of the particular series of controls and depends only on the first and the last control.

\section{Commutativity of increasing controls}

Suppose that we increase one control field while keeping a second constant, allow the system to relax, then increase the second control while keeping the first constant at its new, raised value. As an alternate protocol, suppose that we exchanged the order in which we increased the controls, while still increasing both fields by the same amount. In either case, the system passes through three values of the controls, with the initial and final values the same in the two protocols. Further, in both protocols, the second value of the controls lies between the first and the third.

Thus, according to the equivalence of bounded paths obtained above, both protocols result in a final state that depends only on the initial and final control fields. Since these initial and final controls are the same, the resulting final state is also the same. Thus, increasing controls commute with one another, in the sense that it doesn't matter in which order they are applied. By the same logic, two decreases to the controls also commute with one another.

\section{Noncommutativity of increasing and decreasing controls}

Suppose that one  increases one control field, allows the system to relax and then decreases a second control field.  The controls are now no longer bounded, since some final controls lie strictly above their initial value and some lie strictly below it. As such, the equivalence of bounded paths does not apply to this scenario, and the resulting final states can be different under the two protocols, as shown in the main text via an example simulation.

In fact, the state resulting from the protocol in which the increase in a control field is applied first must lie above (in the partial ordering sense) the state resulting from the protocol in which a control field is decreased first. The reason is that the increase in a control field can flip some spins up, which can increase the local fields $h_i$ to which other spins are subject. If those increases are enough, even once the second control  is turned down, some local fields $h_i$ will remain positive even if they would have attained negative values under the alternate protocol. As such, the corresponding spins can be flipped up under the first protocol but down under the second, which can in turn lead to other spins being flipped up that would otherwise have been flipped down.

\section{Return-point memory}

Consider a protocol that is bounded in the sense described previously, but with the additional feature that for some intermediate timestep the controls are brought to their final state:

\begin{align}
    c_a^{(1)} \le c_a^{(n)}  \le c_a^{(N)}, \textrm{ for all } n,a, \\
    c_a^{{(n')}}=c_a^{{(N)}}, 1 \le n' \le N.
\end{align}

\noindent Consider the state of the system after time step $n'$. According to our previous analysis, this is simply $f(s,\{c_a^{(1)},c_a^{(n')}\})=f(s,\{c_a^{(1)},c_a^{(N)}\})$, which is also the state after time step $N$. That is, despite undergoing complicated evolution through different states that is difficult to predict, the system is guaranteed to return to its intermediate state when the controls are returned to their previous value. 

Note that this does not imply that the state of the system is only a function of the initial state and the applied fields. As shown in the main text through an example, when controls are not bounded in this way two different protocols that end in the same final controls will often result in different states, and many different stable states can be attained at the same value of the control fields.

 %While the vast majority of pairs of microstates are not ordered in this sense, the states obey a \emph{no-passing condition}, in which two ordered states maintain their order when subjected to the same fields. This type of condition was first identified for sliding charge-density waves~\cite{middleton1992asymptotic} and later extended to spin systems~\cite{sethna1993}.

\nocite{*}
\bibliography{citations}
% \clearpage
% \section{Supplementary Material}
% \subfile{supplementary_material_content}